\documentclass[preprint2]{aastex}

\begin{document}

\def\SN{signal-to-noise ratio}
\def\NH{$N_{\rm H}$}
\def\mass{$M_{\odot}$}
\def\deg{^{\circ}}
\def\kps{km s$^{-1}$}
\def\etal{{\it et al.}}
\def\mperyr{$M_{\odot }$ yr$^{-1}$}
\newcommand{\eex}[1]{\times 10^{#1}}
\newcommand{\ARAA}[2]{ARAA\rm, #1, #2}
\newcommand{\ApJ}[2]{ApJ\rm, #1, #2}
\newcommand{\ApJL}[2]{ApJ\rm, #1, L#2}
\newcommand{\ApJSS}[2]{ApJS\rm, #1, #2}
\newcommand{\AandA}[2]{A\&A\rm, #1, #2}
\newcommand{\AJ}[2]{AJ\rm, #1, #2}
\newcommand{\BAAS}[2]{BAAS\rm, #1, #2}
\newcommand{\ASP}[2]{ASP Conf. Ser.\rm, #1, #2}
\newcommand{\JCP}[2]{J. Comp. Phys.\rm, #1, #2}
\newcommand{\MNRAS}[2]{MNRAS\rm, #1, #2}
\newcommand{\N}[2]{Nature\rm, #1, #2}
\newcommand{\PASJ}[2]{PASJ\rm, #1, #2}
\newcommand{\PASP}[2]{PASP\rm, #1, #2}
\newcommand{\RPP}[2]{Rep. Prog. Phys.\rm, #1, #2}

\title{The Detection of a Cooling Flow Elliptical Galaxy from \ion{O}{6} Emission}
\author{Joel N.\ Bregman, Eric D.\ Miller, and Jimmy A.\ Irwin}
\affil{Dept.\ of Astronomy, University of Michigan, Ann Arbor, MI 48109-1090}

\begin{abstract}

Cooling flow models for the hot gas in elliptical galaxies predict that gas
is cooling at a rate of $\sim 1$ \mperyr, yet there is little evidence for
this phenomenon beyond the X-ray waveband. If hot gas is cooling, it will
pass through the $3\eex{5}$ K regime and radiate in the \ion{O}{6} 
$\lambda\lambda$1032,1038
ultraviolet lines, which can be detected with the \textit{Far Ultraviolet
Spectroscopic Explorer} (\textit{FUSE}) and here we report on \textit{FUSE}
observations of the X-ray bright early-type galaxies NGC 1404 and NGC 4636.
In NGC 1404, the \ion{O}{6} doublet is not detected, implying a cooling rate
$ < 0.3$ \mperyr, which is below the predicted values from the cooling flow
model of 0.4--0.9 \mperyr.
In NGC 4636, both \ion{O}{6} lines are clearly
detected, indicating a cooling rate of $0.43 \pm 0.06$ \mperyr, which falls
within the range of values from the cooling flow prediction, 
0.36--2.3 \mperyr and is closest to the model where the production of
the cooled gas is distributed through the galaxy.  The emission line widths,
$44 \pm 15$ \kps, are close to the Doppler broadening value (30 \kps),
indicating that the flow is quiescent rather than turbulent, and that the
flow velocity is $<$ 30 \kps.

\end{abstract}

\keywords{galaxies: individual (NGC 1404, NGC 4636) --- galaxies: ISM}

\section{Introduction}

In the standard model for cooling flow ellipticals, the gas shed by stars
during normal stellar evolution becomes thermalized, colliding with ejecta
from other stars and with the ISM. This process heats the gas to 
$10^{6.5}$--$10^{7}$ K, and if there were no additional heating or cooling,
the gas would be bound to the galaxy and have the same spatial distribution
as the stars. However, radiative cooling drains energy from the gas most
rapidly at small radii, causing a loss of buoyancy and a subsequent inflow
of gas. The rate at which gas cools and flows inward $(\dot{M})$ is
proportional to two observed quantities, the energy loss rate $(L_{X})$
divided by the thermal energy per gram $(3kT_{X}/\mu m_{p})$, or 
$\dot{M} \propto L_{X}/T_{X}$, typically 0.03--3 \mperyr\ for ellipticals.

Although cooling flow models do an adequate job in explaining the X-ray
data, there is little confirmation for the consequences of the models.  In
particular, the cooling flow will produce cooled gas, yet gas at 
$\lesssim 10^{4}$ K is not seen in abundance in these systems (Roberts et
al. 1991), suggesting that the gas is turned into stars. However, if stars
are produced with a normal initial mass function, young blue stars would
have been detected if the rate was above 0.01 \mperyr\ (O'Connell 1999).
Therefore, either star formation in these systems does not produce
many stars above 1 \mass, or the basic model is wrong.  Most of the
modifications to the basic model involve some sort of heating mechanism to
balance the radiative cooling and prevent cool gas from being produced.
The critical issue is whether the gas is losing its thermal energy and a
simple test is to search for the emission lines that are produced as gas
cools through intermediate temperatures.

The most powerful diagnostic line for such a test is the \ion{O}{6}
doublet, which is produced as the gas passes through the $3\eex{5}$ K
range, at which point the gas has lost 90--95\% of its original thermal
energy, so it has effectively cooled.  As gas cools through the 
2--4$\eex{5}$ K range, the cooling is carried by a single ionization state
of a single element, \ion{O}{6}, and its doublet lies at 
$\lambda\lambda$1032,1038, now accessible with the \textit{Far Ultraviolet
Spectroscopic Explorer} (\textit{FUSE}). The relationship between line
luminosity and cooling rate ($\dot{M}$) is insensitive to the metallicity
of the gas or to whether the gas is out of equilibrium (Edgar and Chevalier
1986). Consequently, the luminosity of the $\lambda\lambda$1032,1038 lines
is a direct measure of $\dot{M}$.  We have a program using \textit{FUSE}
(Moos et al. 2000) to observe cooling flow ellipticals and here we report
on the first two observation, which are of the classic cooling flow
galaxies NGC 4636 and NGC 1404.

\section{Observations}

Out of a more extensive program, NGC 1404 and NGC 4636 are the first
two galaxies to be observed, and they are X-ray
bright, so their emission is dominated by hot gas (e.g.,
Canizares, Fabbiano, and Trinchieri 1987).  Furthermore, the cooling time
of the X-ray emitting gas is far less than a Hubble time, 
making these classic cooling flow galaxies.  There
are important differences between these two galaxies in their optical
luminosity as well as their X-ray to optical luminosity ratio, leading to
different values for the mass cooling rate $\dot{M}$. The galaxy NGC 1404
is in the Fornax cluster and for a distance of 28.4 Mpc 
($H_o$ = 50 \kps\ Mpc$^{-1}$; the conclusions
are distant independent), $\log{L_B} = 10.74$, $\log{L_X} = 41.27$ and
$T_X$ = 0.56 keV (Faber et al. 1989; Brown and Bregman 2000), while NGC
4636 is in the outskirts of the Virgo cluster with a distance of 26.7 Mpc
and with properties $\log{L_B} = 10.96$, $\log{L_X} = 41.81$ and 
$T_X$ = 0.72 keV.

\begin{figure}
\plotone{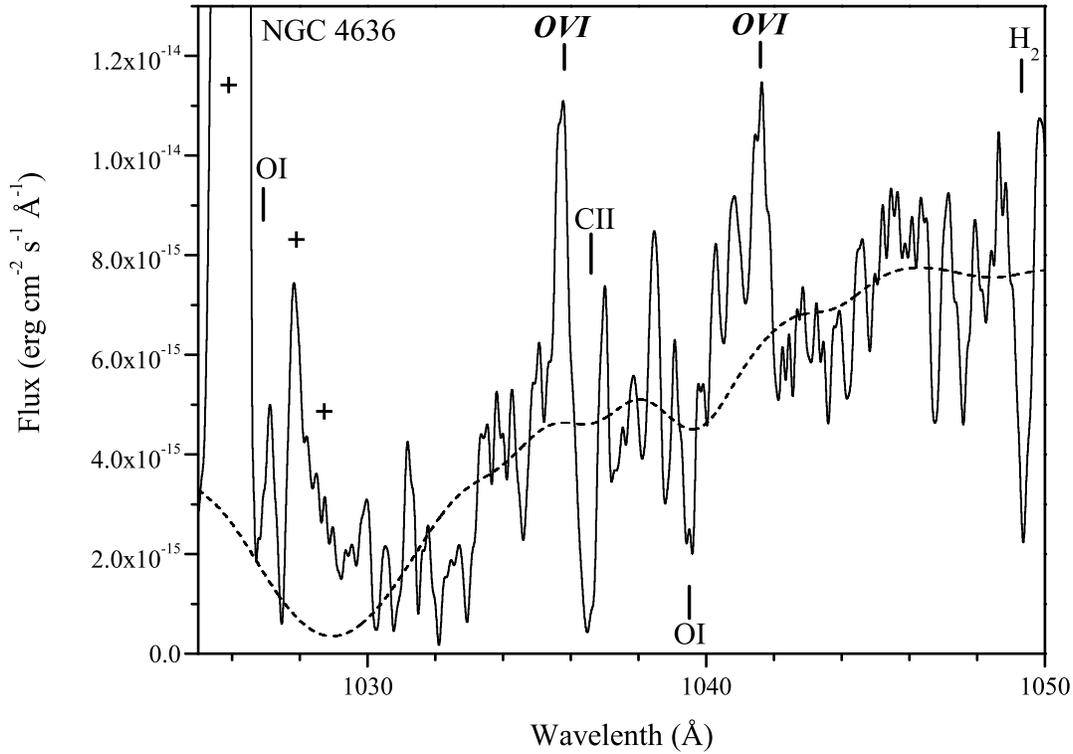}
\caption{
The FUSE spectrum of NGC 4636 in the 1020--1050 \AA\ region shows
airglow lines (notably Ly$\beta$; denoted by +), several Galactic
absorption lines, the strongest being \protect\ion{C}{2} (1036.34 \AA) and
\protect\ion{O}{1} (1026.47 \AA, 1039.23 \AA), and possibly
\protect\ion{O}{6} (1032 \AA), as well as the redshifted \protect\ion{O}{6}
emission lines of from NGC 4636 (in italics).  The dashed line is the
expected stellar continuum from NGC 4636.  The spectrum has been smoothed
to a velocity resolution of 100 \protect\kps\ for display purposes,
although from the full resolution data, the \protect\ion{O}{6} emission
lines have a FWHM of 44 \protect\kps, close to the thermal width of 30
\protect\kps.  \label{fig:4636}
}
\end{figure}

The observations for NGC 1404 were obtained in 10 December 1999, with an
exposure time of 7.49 ksec, while the data for NGC 4636 were obtained in 23
May 2000 with an exposure time of 6.46 ksec. In both cases, the large
aperture was used, which is 30\arcsec\ square, leading to a velocity 
resolution of 100 \kps; there is no effective spatial 
information perpendicular to the dispersion axis.  The data streams did not
contain periods of bursts or other detrimental events, so editing of the
event file was not necessary.  The data were processed with the pipeline
program during January 2001, including an optimal extraction algorithm, which
uses a local background and employs information about the shape of the 
spectrum perpendicular to direction of the spectrum.
Four different detector segments cover the spectral region of interest, two
LiF segments and two SiC segments, but the LiF1a channel had the highest
$S/N$ and was used in most of the analysis.  Addition of the other channels
generally reduced the $S/N$ and led to spurious features in the spectrum.
The 30\arcsec\ aperture was centered on
the optical centers of the galaxies and the resulting spectra around the
regions of the \ion{O}{6} emission in NGC 4636 and NGC 1404 are show in
Figures \ref{fig:4636} and \ref{fig:1404}.

These spectra have features not associated with the elliptical
galaxies, such as airglow lines and Galactic absorption lines.
These lines of sight are out of the plane of the Milky Way, where the
\ion{H}{1} columns are $1.36 \eex{20}$ cm$^{-2}$ (NGC 1404) and 
$1.63 \eex{20}$ cm$^{-2}$ (NGC 4636), and at these low values, the H$_{2}$
absorption lines are not expected to be large. 
The strongest Galactic atomic line in the region is that of \ion{C}{2} at
1036.34 \AA, followed by \ion{O}{1} lines at 1026.47 \AA\ and 1039.23 \AA.
A few other lines fall in this window, the ground-state doublet of 
\ion{O}{6} $\lambda\lambda$1031.95,1037.63 and the excited metastable line
\ion{C}{2}$^{\ast}$ $\lambda$1037.02, and although Galactic absorption by
these lines is seen in a number of other sightlines
through the galaxy (e.g., Sembach et al. 2000), they are weaker.
These two \ion{O}{6} lines are the ones predicted to be present in cooling
flows in emission, and for optically thin conditions, the 
\ion{O}{6} $\lambda$1032 line will be twice as strong as the 
\ion{O}{6} $\lambda$1038 line.

\subsection{NGC 1404}

\begin{figure}
\plotone{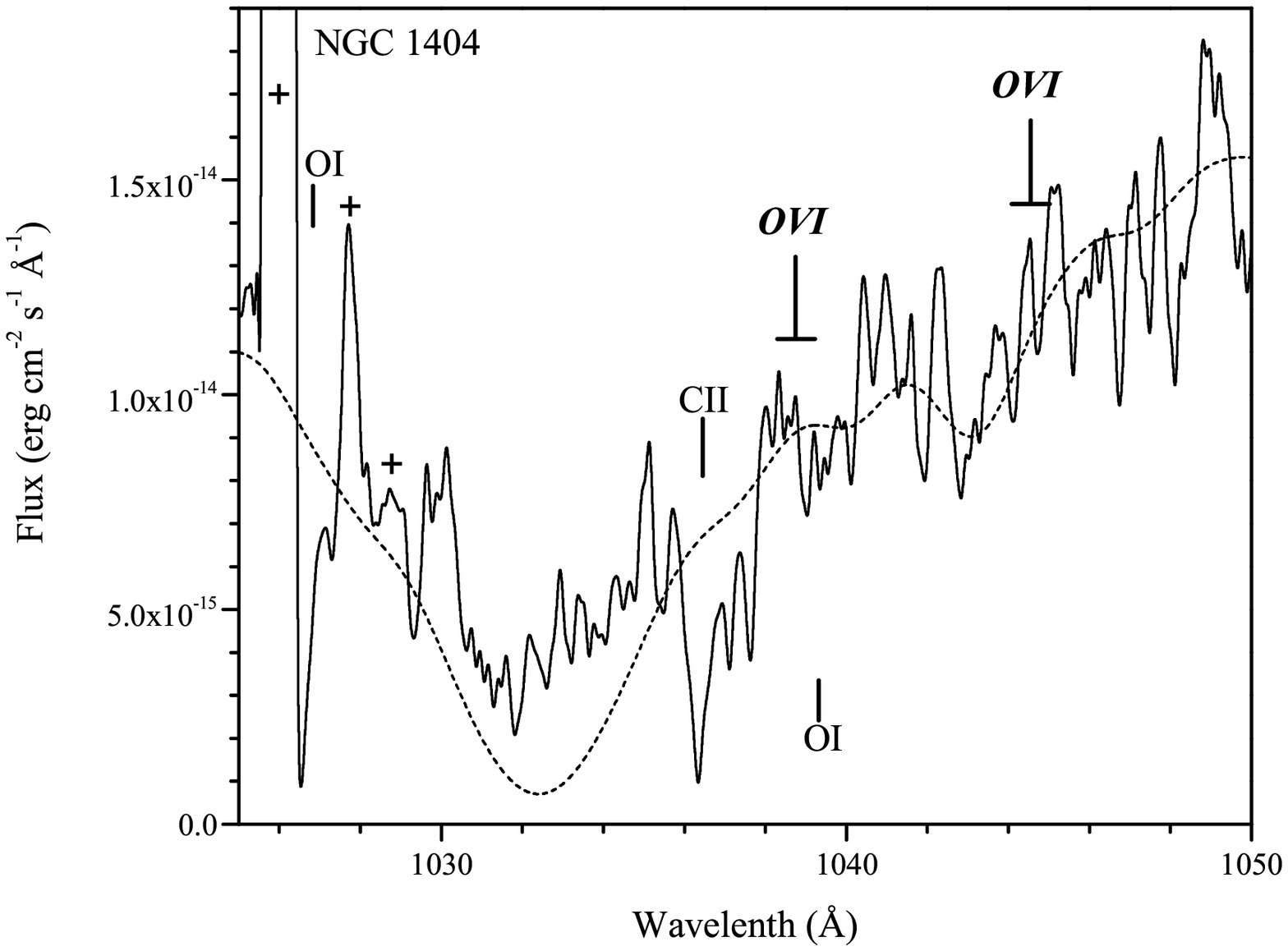}
\caption{
The FUSE spectrum of NGC 1404, similar to that in Figure \ref{fig:4636},
except there is no apparent detection of the redshifted \ion{O}{6} emission
lines from NGC 1404.  \label{fig:1404}
}
\end{figure}

Of the four segments, only the LiF1a had a count rate above the background
rate and it produces the highest $S/N$ spectrum in the 1025--1050 \AA\ range
(Fig.\ \ref{fig:1404}). 
The spectrum in the range 1025--1050 \AA\ range shows a stellar
continuum upon which terrestrial airglow lines plus a few Galactic
absorption features are present (Fig.\ \ref{fig:1404}): the strong line of 
\ion{C}{2} $\lambda$1036.34 at zero velocity; an \ion{O}{1} $\lambda$1026.47 
line adjacent to the Ly$\beta$ airglow feature; and Ly$\beta$ absorption, although 
it is filled in by the Ly$\beta$ airglow line plus two other airglow lines at
1027--1028 \AA. Given the modest intensity of the airglow line at 1025.5 \AA, 
the other airglow lines in this spectral region should be unimportant, and they
are generally not visible (e.g., the \ion{O}{1} $\lambda$1039.2 line is
absent).  The strongest H$_{2}$ lines in this and other spectral
regions are not visible, so the Galactic $N(H_{2}) < 10^{18}$ cm$^{-2}$.
There are no interstellar atomic or molecular absorption lines at the
redshift of NGC 1404.

The stellar continuum has one dominant feature in Figure \ref{fig:1404},
Ly$\beta$ absorption, plus a number of minor absorption lines. In an effort
to understand which features are stellar and which might be interstellar,
we can compare it to an approriate stellar model. As discussed by O'Connell
(1999), the continuum in this region is produced by low mass stars that
either populate the extreme horizontal branch or have evolved away from
it.  Models that include these various stellar contributions were
calculated by Brown et al.\ (1997) for comparison with HUT data, with
atmospheric metallicities of $0.1 Z_{\odot}$ leading to the best fits.
These low atmospheric abundances probably reflect diffusion processes since
the mean metallicities of the stars in these systems are near-solar. The
model of Brown et al.\ (1997) is at 3 \AA\ resolution, which is too low for
comparison with our data, so instead we use the average of two stars of
different gravity ($g = 4, 5.5$), as these are representative of the range of
gravities for the stars that contribute to the flux in this wavelength
region (Dorman, Rood, and O'Connell 1993); we used a surface metallicity of
$0.1 Z_{\odot}$ and $T = 26,000$ K. Also, this model was convolved with the
velocity dispersion of the galaxy (Fig.\ \ref{fig:1404}).

The expected \ion{O}{6} emission lines from NGC 1404 would be shifted to
$\lambda\lambda$1038.6,1044.3 \AA (1947 \kps), and relative to the continuum,
there is no evidence of emission from either line. To estimate
the upper limit to the line strength, we assume that the line width is
characteristic of the velocity dispersion of the system,  
$\Delta\lambda = 1.0$ \AA\ (300 \kps).  Conservative 3$\sigma$ upper limits for the
lines are 
$F(\lambda 1032) < 3\eex{-15}$ erg cm$^{-2}$ s$^{-1}$, and 
$F(\lambda 1038) < 1.4\eex{-15}$ erg cm$^{-2}$ s$^{-1}$.
The extinction in this region is very low, estimated to be $A_V$ = 0.00-0.05,
so the average would lead to a extinction correction of a factor of 
1.08 at 1035 \AA.

\subsection{NGC 4636}

The spectrum of NGC
4636 contains two Galactic features seen in the spectrum of NGC 1404, the
\ion{C}{2} $\lambda$1036.34 and the \ion{O}{1} $\lambda$1039.23 feature
(Fig.\ \ref{fig:4636}), although they are shifted by 0.24 \AA\ and 
0.30 \AA\ respectively from their rest wavelengths (about $+70$ \kps).
These Galactic absorption lines are expected to occur near 0 \kps, yet
there is no line at this location with the instrumental width.  Also there
is no known Galactic \ion{H}{1} feature of any importance at $+70$ \kps,
which would need to dominate the neutral column in this direction to cause
the observed results.  Instead, it appears that there is a spectral shift
of about 0.3 \AA, and a similar shift is seen in the Ly$\beta$ airglow
line, which is redshifted by 0.40 \AA.  Consequently, in our analysis, we
apply a shift to the data of -0.3 \AA\ ($-87$ \kps).  In addition to the
atomic lines, there is modest absorption by H$_{2}$, although only the
strongest lines are detected, such as the 1049.7 \AA\ line. Based on the
strength of this and other Galactic H$_{2}$ (e.g., in the 1062.5--1064 
\AA\ region), we estimate that $N(H_{2}) \approx 10^{18.5}$ cm$^{-2}$, typical
of lines of sight with this $N$(\ion{H}{1}) value.

Aside from the absorption lines and airglow lines, the two strongest
features are emission lines that correspond to the wavelengths at which the
\ion{O}{6} doublet would occur near the redshift of NGC 4636; the line
centers are $1018 \pm 20$ \kps\ ($\lambda$1032 line) and $1061 \pm 20$ 
\kps\ ($\lambda$1038 line) compared to the galaxy redshift of 
$938 \pm 4$ \kps\ (from NED).  The separation of the line peaks is, to
within the errors, consistent with the separation of the two lines in the
\ion{O}{6}, but the mean velocity of the lines is 100 \kps\ greater than 
that expected.  This difference may result from wavelength calibration errors 
since in a previous version of the pipeline processing, the wavelength of the
\ion{O}{6} was consistent with optical redshift for NGC 4636.

The two lines are single-peaked and fairly narrow.  The FWHM of the lines are
about $0.15 \pm 0.05$ \AA\ (44 \kps), with the error depending upon where
one defines the continuum.  The instrumental resolution depends upon the
degree to which the aperture is illuminated.  The airglow lines, which fill
the aperture uniformly, have a FWHM of 0.34 \AA\ (100 \kps), while a point
source would have a FWHM of about 0.05 \AA\ (15 \kps).
Therefore, the emission lines from NGC 4636 must not fill the aperture uniformly,
although we cannot determine whether the line width is due to an instrumental 
limitation or to Doppler broadening.
The line fluxes are 
$F(\lambda 1032) = 3.0 \pm 0.6 \eex{-15}$ erg cm$^{-2}$ s$^{-1}$, and 
$F(\lambda 1038) = 2.4 \pm 0.6 \eex{-15}$ erg cm$^{-2}$ s$^{-1}$, with the
primary error in the line fluxes resulting from the uncertain placement of the continuum.
The line ratio, 1.25 $\pm$ 0.32, is 2.3$\sigma$ from the optically thin 
value of 2.0, but given the uncertainty in the model stellar continuum,
we do not consider this difference particulary significant.
The extinction along this sightline is $A_V$ = 0.04--0.09, depending upon the reddening
model used (Burstein and Heiles 1982, or Schlegel, Finkbeiner, and Davis 1998), so
we use the average of the two values, along with a value of $A(1035~\rm{\AA})/A_V$ = 4.0 
(Cardelli et al. 1989), which leads to an exctinction-corrected value of 
$F(\lambda 1032+\lambda 1038) = 6.8 \pm 0.9 \eex{-15}$ erg cm$^{-2}$ s$^{-1}$.

\section{Interpretation and Discussion}

The conversion between the \ion{O}{6} line luminosity and cooling rate
depends only weakly on the details of the physical situation, provided that
the initial temperature is significantly higher than the temperature at
which the ion is most abundant.  This situation occurs in these galaxies,
where the ambient temperature of the hot X-ray emitting gas is 
6--9$\eex{6}$ K and the ionization fraction in \ion{O}{6} is negligible.
The cooling of the gas will proceed isobarically if the cooling time is
long compared to the sound-crossing time of the region, eventually becoming
isochoric as the gas cools and the relative size of the timescales is
reversed.  If gaseous cooling occurs in regions smaller than 1 kpc, the
transition from the isobaric to isochoric case occurs at 
$T \lesssim 4\eex{5}$ K (Edgar and Chevalier 1986), and the cooling may be
entirely isobaric as it applies to \ion{O}{6}.  Unfortunately, we do not at
this time have a good contraint on the size of the cooling region.
For the pure
isobaric case, Edgar and Chevalier (1986) show that 
$L(\lambda 1032) = 1.1 \eex{39} \dot{M}$ erg s$^{-1}$ ($\dot{M}$ is in 
\mperyr) and for the pure isochoric case, 
$L(\lambda 1032) = 0.7 \eex{39} \dot{M}$ erg s$^{-1}$ (the $\lambda$1038
line is half of the strength of the $\lambda$1032 line).  In this
paper, we will use a conversion that is halfway between the isochoric and
isobaric cases, $L(\lambda 1032) = 0.9\eex{39} \dot{M}$ erg s$^{-1}$, which
introduces only a 20\% uncertainty in the model; the resulting values of
$\dot{M}$ are given in Table \ref{tab}.

The critical test is to compare these values of $\dot{M}$ with those
derived from X-ray measurements. The cooling rate $\dot{M}$ is the rate at
which the thermal energy of the system is being drained away by radiative
losses, so it is the ratio of the net cooling rate divided by the specific
thermal energy content, or $\dot{M}=L_{net}/E$, where the specific thermal
energy is $E=3kT/\mu m_{p}$, $\mu$ is the mean molecular weight for the hot
gas ($\mu =0.63$) and $m_{p}$ is the usual proton mass.  The net energy
loss rate $L_{net}$ equals the radiative loss rate (the bolometric X-ray
luminosity) minus heating sources such as supernovae or gravitational
compression as the gas falls inward.  The amount of gravitational
compressional heating depends upon whether the gas cools close to where it
entered the flow or whether it flows into the central region 
($r\lesssim r_{core}$) before losing most of its thermal energy.  The gas
will become cool only in the central region unless thermal instabilities
can grow throughout the flow, but the best calculations on this matter
indicate that most linear perturbations will not grow rapidly enough to be
effective (Balbus 1988).  However, models without distributed mass drop-out
lead to an X-ray surface brightness profile that is too sharply peaked in
the center, inconsistent with observations.  Consequently, most models
include mass drop-out as a function of radius and the usual formulation is
that the rate of mass drop-out is inversely proportional to the cooling
time, or $\dot{\rho}=q\rho/t_{c}$, where $t_{c}$ is the isobaric cooling
time (Sarazin and Ashe 1989).

Without mass drop-out, $q=0$, and a moderately efficient rate of drop-out
would be given by $q=1$, so we consider both cases.  We use the model from
Sarazin and Ashe (1989), where $L_B = 1\eex{11}$ $L_{\odot}$, and expressing
$L_X$ in units of $10^{41}$ erg s$^{-1}$ and $T_X$ in keV, we obtain
$\dot{M}=0.26 L_X T_{X}^{-1}$ \mperyr\ for $q=0$, and 
$\dot{M}=0.40 L_X T_X^{-1}$ \mperyr\ for $q=1$ (this is for the
absorption-corrected X-ray luminosity in the 0.5--2.0 keV band).  In the
$q=1$ case, the mass drop-out occurs throughout the galaxy and since our
30\arcsec\ square aperture does not encompass the entire galaxy, we must
determine $\dot{M}$ within the aperture.  For our two galaxies, the square
aperture encloses the emission from the galaxy within about 2.2 kpc (NGC
4636) and 2.4 kpc (NGC 1404) from the center, and in the $q=1$ model, this
would enclose about half of the total $\dot{M}$.  After applying this
factor to the $q=1$ case, the resulting $\dot{M}$ is similar to the $q=0$
case, with $\dot{M}=0.20 L_X T_X^{-1}$ \mperyr.  We use the values of
$L_{X}$ and $T_{X}$ given in Brown and Bregman (2000) for these galaxies to
calculate values of $\dot{M}$ for the $q=0$ case and for the $q=1$ case for
the cooling only within the aperture (denoted $\dot{M}_{0}$, $\dot{M}_{1}$
and given in Table \ref{tab}).

An alternative method of determining the expected value of $\dot{M}$ is to
use the X-ray observations directly by noting that 
$\dot{\rho}=q\rho/t_{c}\propto \rho^{2}/\Lambda (T_{X})$, but since the
cooling function $\Lambda (T_{X})$ changes far less than $\rho$ in these
galaxies, a good approximation is $\dot{\rho} \propto \rho^{2}$, and the
amount of cooling gas in the aperture is the product of the line integral
of $\dot{\rho}$ through the galaxy, or $\dot{M} \propto A\int \rho^{2}dl$,
where $A$ is the area of the aperture.  The quantity $\int \rho ^{2}dl$ is
the emission measure, to which the X-ray surface brightness is
proportional, so the fractional amount of $\dot{M}$ within the aperture is
just $\dot{M}_{A}/\dot{M}_{Tot}=L_{X,A}/L_{X,Tot}$.  We have calculated the
quantity $L_{X,A}/L_{X,Tot}$ directly from the ROSAT PSPC data for these
systems (using the backgrounds discussed by Brown and Bregman 2000) and
obtain the fractional cooling within the \textit{FUSE} aperture to be 0.083
(NGC 4636) and 0.25 (NGC 1404).  This leads to values of $\dot{M}$ within
the apertures of 0.36 \mperyr\ (NGC 4636) and 0.40 \mperyr\ (NGC 1404),
which is the product of $L_{X,A}/L_{X,Tot}$ and the value of $\dot{M}$
given above for the $q=1$ case (given as $\dot{M}_{2}$ in Table \ref{tab}).
For NGC 1404, this value of the cooling rate in the aperture is within a
factor of two of that derived entirely from the Sarazin and Ashe model.
However, in the case of NGC 4636, the two values differ by up to a factor
of six.  Probably, this difference arises because NGC 4636 is more extended
than the average galaxy of similar optical luminosity.  Whereas NGC 4636
and NGC 1404 are similar in both optical luminosity and distance, NGC 4636
has a value for the half-light radius, $r_{e}$, that is nearly four times
greater than for NGC 1404 (101\arcsec\ compared to 27\arcsec).
Consequently, the core radius used in the model by Sarazin and Ashe (1989)
may be too small in the case of NGC 4636 and this would lead to an
overestimation of the fractional amount of $\dot{M}$ within the aperture
for the $q=1$ case; the values of $\dot{M}$  are not affected for the $q=0$
case.  The different values for $\dot{M}$ from the X-ray data may be
regarded as the expected range in the prediction, given the uncertainties
in the models.

\section{Conclusions}

For NGC 4636, the value of $\dot{M}$ from the \ion{O}{6} measurement lies
within the predicted range of $\dot{M}$ from the X-ray data, lending strong
support to the cooling flow model.  Furthermore, the \ion{O}{6} result
indicates a cooling rate significantly below that predicted from the $q=0$
model, a discrepancy that is removed if we assume that the cooling is
distributed throughout the galaxy, which also was predicted from the
application of the cooling flow models to the X-ray surface brightness
distribution.  Since $\dot{M} ($\ion{O}{6}$) \approx \dot{M}_{2}$, the mass
drop-out is consistent with that obtained by assuming that
$\dot{M}(<r) \propto L_X(<r)$, although more detailed models providing
determinations of $\dot{M}$ should be calculated for NGC 4636, such as
along the lines of Bringhenti and Mathews (1996, 1999), who include a more
accurate galaxy model plus important effects of accretion within the galaxy
group.

In NGC 4636, the \ion{O}{6} line width of $44 \pm 15$ \kps\ is similar to
the thermal Doppler width of 29 \kps\ and is less than either the
sound speed in the gas, 77 \kps, or the velocities of the $10^{4}$ K
optical emission line gas of 200 \kps\ (Caon et al. 2000).
The narrow line width has several implications, the first being that the radial 
velocities within the cooling flow are less than 30 \kps\ (3 $\sigma$ upper limit).
This is consistent with the cooling flow calculations, which predict typical
flow velocities of $\sim$ 10 \kps\ (e.g., Sarazin and Ashe 1989).  
Furthermore, the environment of the \ion{O}{6} gas must be quiescent, 
also consistent with models.

In contrast to NGC 4636, for NGC 1404 the predictions of the cooling flow model
are not confirmed in the \ion{O}{6} data, where our upper limit is below the
lowest X-ray prediction for $\dot{M}$.  The most likely resolution of this
discrepancy is that the net radiative loss rate is not given by the X-ray
luminosity, which will occur if there is an additional source of heating, such
as from supernovae or a central AGN.  There is no evidence for a central radio
source, as there is only an upper limit to the radio continuum emission of 0.7
mJy (6 cm; Sadler, Jenkins, and Kotanyi 1989).

In the future, we will observe additional galaxies to determine whether the
cooling flow model is generally applicable. Also, we will obtain additional
observations of NGC 4636 in off-central locations, in order to determine
the radial distribution of the cooling material, which is vital to modelers
and to the determination of the location of the deposition of cooled gas
that may produce new stars.

We would like to thank the considerable efforts of the \textit{FUSE} staff
in carrying out the observations and assisting us with the data processing;
we received invaluable assistance from B-G Anderson, G. Sonneborne, W.
Oegerle, J. Kruk, K. Sembach, and W. Blair. Also, we wish to acknowledge
the insights and advice provided by W.G. Mathews, R. Edgar, R.
O'Connell, T. Brown, and B. McNamara. Support for this program was provided by NASA
through grant NAG5-9021.

\onecolumn

\begin{deluxetable}{ccccc}
\tablewidth{280pt}
\tablecaption{Cooling Rates from X-Ray and \ion{O}{6} Observations.\tablenotemark{a} \label{tab}}
\tablehead{
\colhead{Galaxy Name} &
\colhead{$\dot{M}$(\ion{O}{6})} &
\colhead{$\dot{M}_{0}$} &
\colhead{$\dot{M}_{1}$} &
\colhead{$\dot{M}_{2}$} 
}
\startdata
NGC 1404 & $<$ 0.3           & 0.86 & 0.66 & 0.40 \\
NGC 4636 & 0.43 $\pm$ 0.06 & 2.3  & 1.8  & 0.36 \\
\enddata
\tablenotetext{a}{Units of $\dot{M}$ are \mperyr.}
\end{deluxetable}

\begin{references}

\reference{} Balbus, S.A. 1988, ApJ, 328, 395

\reference{} Brighenti, F., and Mathews, W.G. 1996, ApJ, 470, 747

\reference{} Brighenti, F., and Mathews, W.G. 1999, ApJ, 512, 65

\reference{} Brown, B.A., and Bregman, J.N. 2000, ApJ, 539, 292

\reference{} Brown, T.M., Ferguson, H.C., Davidsen, A.F., and Dorman, B.
1997, ApJ, 482, 685

\reference{} Burstein, D., and Heiles, C. 1982, AJ, 87, 1165.

\reference{} Caon, N., Macchetto, D., and Pastoriza, M. 2000, ApJS, 127, 39

\reference{} Canizares, C.R., Fabbiano, G., and Trinchieri, G. 1987, ApJ,
312, 503

\reference{} Cardelli, J.A., Clayton, G.C., and Mathis, J.S. 1989, ApJ, 345, 245

\reference{} Dorman, B., Rood, R.T., and O'Connell, R.W. 1993, ApJ, 419, 596

\reference{} Edgar, R.J., and Chevalier, R.A. 1986, 310, L27

\reference{} Faber, S.M. et al. 1989, ApJS, 69, 763

\reference{} Moos, H.W., et al. 2000, ApJL, 538, L1

\reference{} O'Connell, R.W. 1999, ARAA, 37, 603

\reference{} Roberts, M.S., Hogg, D.E., Bregman, J.N., Forman, W.R., and
Jones, C. 1991, ApJS, 75, 751

\reference{} Sadler, E.M., Jenkins, C.R., and Kotanyi, C.G. 1989, MNRAS, 240, 591

\reference{} Sarazin, C.L. and Ashe, G.A. 1989, ApJ, 345, 22

\reference{} Schlegel, D.J., Finkbeiner, D.P., and Davis, M. 1998, ApJ, 500, 525.

\reference{} Sembach, K.R., et al. 2000, ApJL, 538, L31

\end{references}
\end{document}